\documentclass[twocolumn,aps,floatfix,amsmath,amssymb,superscriptaddress,pre]{revtex4}
\usepackage{graphicx}

\newcommand{\be}{\begin{equation}}
\newcommand{\ee}{\end{equation}}
\newcommand{\beqn}{\begin{eqnarray}}
\newcommand{\eeqn}{\end{eqnarray}}

\begin{document}

\title{Hybrid multilane models for highway traffic}

\author{Taiyi Zhang}
\author{Yu-Cheng Lin}
\email{yc.lin@nccu.edu.tw}
\affiliation{Graduate Institute of Applied Physics, National Chengchi University, Taipei 11605, Taiwan}

\begin{abstract}
We study effects of lane changing rules on multilane highway traffic using the
Nagel-Schreckenberg cellular automaton model with different schemes for
combining driving lanes (lanes used by default) and overtaking lanes.  Three
schemes are considered: a symmetric model, in which all lanes are driving
lanes; an asymmetric model, in which the right lane is a driving lane and the
other lanes are overtaking lanes; a hybrid model, in which the leftmost lane is
an overtaking lane and all the other lanes are driving lanes.  In a driving lane
vehicles follow symmetric rules for lane changes to the left and to the right,
while in an overtaking lane vehicles follow asymmetric lane changing rules.  We
test these schemes for three- and four-lane traffic mixed with some low-speed
vehicles (having a lower maximum speed) in a closed system with periodic
boundary conditions as well as in an open system with one open lane.  Our
results show that the asymmetric model, which reflects the ''Keep Right Unless
Overtaking'' rule, is more efficient than the other two models.  An extensible
software package developed for this study is free available. 
\end{abstract}

\date{\today}

\maketitle

\section{Introduction}
\label{sec:intro}

Transport phenomena arise in a wide variety of many-particle systems, ranging
from vehicular traffic to physical or biological systems, such as fluid flow,
molecular motor transport and more.  Among all theoretical transport models,
the totally asymmetric simple exclusion process (TASEP) is the most widely
applied paradigm for transport of interacting particles~\cite{1dTASEP}. The
TASEP consists of particles moving unidirectionally on a one-dimensional
lattice with the constraint that at any given time every lattice site can be
occupied at most by one particle. The Nagel-Schreckenberg (NaSch) cellular
automaton model~\cite{NaSch}, a standard model for the simulation of highway
traffic, can be regarded as an extension of the TASEP in which the maximum
speed of particles (vehicles), acceleration and random deceleration are
introduced to mimic some basic phenomena in highway traffic, such as
spontaneous formation of congestion.  The flexibility of the NaSch cellular
automaton approach allows one to generate various versions of the model to
study different aspects of traffic problems.  While many results are known for
the single-lane model~\cite{1dTASEP,Rev},  there remains much scope for further
study on multilane generalizations.  For example, in multilane traffic models
different types of the lane-changing rules can lead to considerably different
results; therefore, one expects that lane-changing rules in real traffic can
have significant impact on traffic flows.  

A large number of lane-changing rules have been considered in many previous
studies to reproduce various empirically observed
phenomena (see~\cite{Rev,Nagel_PRE,Knospe,Lee,Habel,Su}). 
Instead of proposing new decision-making rules and criteria for
lane changes, here we focus on effects of different arrangements of lanes on
multilane highway traffic.     
We present an extensible {\tt C++} package (available in~\cite{CODES}) for implementing the
multilane NaSch model with a set of tunable parameters and conditions, including
the number of lanes (any positive integer), lane-changing rules, boundary conditions of each
lane, and more.  The significance of this package is its user-definable
settings for the individual lanes.
Using this software package we consider three types of lane arrangements 
implemented on the NaSch model with more than two lanes, corresponding to
three different lane-changing models:
(1) the symmetric model, in which overtaking is allowed on the left and also on the right in all 
lanes; (2) the
asymmetric model, in which the right lane is used by default and
overtaking has to be on the left;
(3) the hybrid model, in which the leftmost
lane is the overtaking lane while the other lanes are treated as in the
symmetric model. The hybrid model describes multilane highway traffic 
observed in many countries where only the lane closest to the median strip
is designed for overtaking and  
overtaking on the right
is not considered to be a driving offense.
By comparing the traffic flow at a fixed number density of vehicles in closed systems
and the average velocity in open systems,  
we demonstrate that for heterogeneous traffic consisting of different types of vehicles 
(fast and slow  vehicles) 
the asymmetric model is more efficient than the other two models.
Here and  throughout the paper we consider ''right-hand traffic'', for ''left-hand traffic'' implemented
such as in the UK and Japan  ''left'' and ''right'' have to be interchanged.

The paper is organized as follows. In Sec.~\ref{sec:program} we briefly
describe the software package and outline the lane changing
criteria as well as observables used in our study; 
in Sec.~\ref{sec:results} after defining the models 
we show results for various quantities in closed and open systems
separately in two subsections. We conclude in Sec.~\ref{sec:summary} with a summary and 
discussion of future prospects.

\section{Program description}
\label{sec:program}

The models underlying the software is the NaSch model,
a cellular automaton with parallel update dynamics, 
i.e. the vehicles are picked up in parallel for updating at each time step. 
Here we recall the definition of the the NaSch model~\cite{NaSch}.
A lane is divided into $L$ cells; at a given time each cell is empty or occupied at most by one 
vehicle with a discrete speed, up to a maximum value: $v\in\{0,1,\cdots,v_{\rm max}\}$.
The update scheme on a single lane
consists of four actions in a time step $t\to t+1$:
\begin{description}
  \item[{\bf A1}]{\it Acceleration:} the velocity of any vehicle that is not at the maximum velocity $v_{\rm max}$ is increased by one unit (measured in cells$/$time step):
\be
	v\to\min\left(v+1, v_{\rm max} \right)\,.
\ee 
  \item[{\bf A2}]{\it Deceleration:} if the distance (in units of cells), $d$, 
between a vehicle we are looking at and the vehicle 
in front of it is smaller than its current velocity $v$, 
the velocity is reduced to $d$ to avoid a collision, i.e.,
\be
        v\to d,\quad{\rm if }\; d<v\,.
\ee  
  \item[{\bf A3}]{\it Random braking:}  The velocity of all vehicles that have $v\ge 1$, is decreased randomly by
one unit with probability $p$. 
  \item[{\bf A4}]{\it Moving:} each vehicle is moved forward to the cell according to its velocity determined 
       in {\bf A1-A3}. 
\end{description}
In the original NaSch model, the randomization parameter $p$ in {\bf A3} is chosen to be a constant;
there is also a generalization, the so-called velocity-dependent randomization (VDR) model~\cite{VDR}, 
in which the probability $p$ depends on the velocity of the vehicle, $p=p(v)$.
We have included these two versions in the package.

For a multilane model, two types of lanes are introduced: overtaking lanes and driving lanes 
(default lanes), 
depending essentially on whether criteria for changing the lane to the left and
to the right are symmetric. Here lane changing is implemented as a sideways move to the neighboring lane,
while forward movement is implemented in single-lane updates ({\bf A1}-{\bf A4}) on each lane 
after possible lane changing of each vehicle is considered, 
i.e. one time step consists of lane changing and single-lane updates.   
In general, there are two types of lane-changing criteria~\cite{Rickert}: incentive criteria and
safety criteria. Following Ref.~\cite{Rickert,Chowdhury}, we include the following incentive criteria in our program:
\begin{description}
 \item[{\bf LC1}] {\it Incentive criterion:} the distance to the vehicle ahead in the same lane is smaller than
   a certain length: $d<\ell$. 
 \item[{\bf LC2}] {\it Incentive criterion:} the distance to the vehicle  ahead in the target lane is larger
 than a certain length: $d_{\rm target}>\ell_{\rm target}$. 
\end{description} 
The safety criteria included in the program are~\cite{Nagel_PRE,Rickert}:
\begin{description}
  \item[{\bf LC3}] {\it Safety criterion:} the target cell is not occupied 
    or there is no ''scheduling conflict'', which happens
    e.g. in a three-lane (sub-)system when a vehicle from the left lane and a vehicle from the right lane 
   are considered to go to the same cell in the middle lane. 
  \item[{\bf LC4}] {\it Safety criterion:} the distance to the vehicle behind in the target lane is larger
   than a certain length: $d^-_{\rm target}>\ell^-_{\rm target}$. 
\end{description}
These four rules are applied to change to the left lane both from an overtaking lane and 
from a driving lane.
In an overtaking lane, which is for overtaking vehicles only, one should return to the right
driving lane after the overtaking maneuver;
thus, only the safety criteria ({\bf LC3} and {\bf LC4}) are considered 
for changing from an overtaking lane to the right lane.
On the other hand, the criteria {\bf LC1}-{\bf LC4} are all required for changing 
from a driving lane to the right lane;
that is, the lane-changing rules for a middle driving lane do not depend on the direction of 
the lane-changing maneuver.
If the criteria with respect to a driving lane both for changing to left and to right are satisfied,
we choose the target lane based on the size of the gap (the distance to the vehicle ahead)
in the left (denoted by $d_l$) and right ($d_r$) lane:
\begin{itemize}
\item Change to left if $d_l>d_r$.
\item Change to right if $d_r>d_l$.    
\item Change to left or right with equal probability if $d_r=d_l$.
\end{itemize}
Note that there have been a variety of lane-change criteria suggested 
in the literature, which can be easily adapted in the code. 
In this paper we use the choices of the parameters $\ell, \ell_{\rm target}, \ell^-_{\rm target}$ 
in the criteria {\bf LC1}, {\bf LC2} and {\bf LC4} as suggested in Ref.~\cite{Chowdhury} and set: 
$\ell=\min(v+1,v_{\rm max})$, $\ell_{\rm target}=d$ and $\ell^-_{\rm target}=v_{\rm max}$.

The program contains a set of adjustable parameters, including the number of
the lanes, the number of cells per lane, the type of each lane (an overtaking
lane or a driving lane), types of vehicles (depending on their maximum
velocities), braking probabilities, lane-changing probabilities, time steps and the number of samples.  
In addition, each lane can be chosen to be closed with periodic boundary conditions or open; 
open lanes and closed lanes can be combined in an arbitrary order into a multilane system.
For an open system, the entry rate is an additional input parameter. 

A number of quantities are measured in the simulation, 
such as the average velocity of vehicles defined as
\be
   \overline{v}=\frac{1}{T}\frac{1}{N}\sum_{t=1}^{T} \sum_{m=1}^N v_m(t)\,,
\ee        
for $N$ vehicles in $T$ time steps, and the flow:
\be
   J=\rho \overline{v}\,,
\ee
where $\rho$ is the number density of vehicles, given by
\be
    \rho=\frac{N}{nL}
\ee
on an $n$-lane road of length $L$.
The observables are collected for the whole system and also for
each individual lane; in the latter case $n=1$ and only vehicles appearing 
in the lane that we are looking at are considered, e.g. the density in the lane $i$:
\be
    \rho_i=\frac{N_i}{L}\,.
\ee
We also distinguish the observables between  different types of vehicles
in heterogeneous traffic.
The results are given as functions of density for a closed system, 
and as functions of entry rate for an open system.
The code also includes a parallel mode for using Message Passing Interface (MPI)  
to distribute different values of the variables to  multiple processors.

\section{Hybrid multilane highway models}
\label{sec:results}

\begin{figure}
\centering
\centerline{\includegraphics[width=0.6\columnwidth]{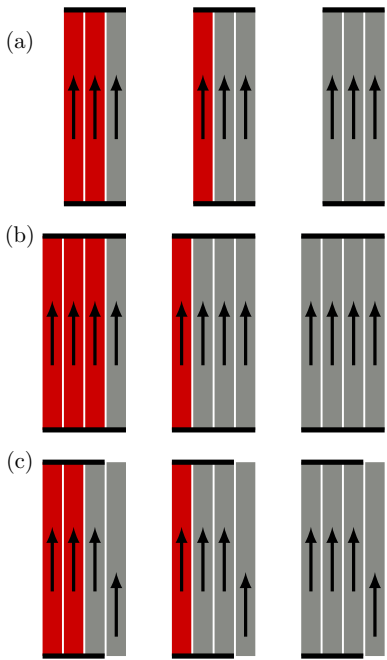}} 
\caption{The models. (a): Three-lane closed systems with periodic boundary conditions (PBC) 
in the traveling direction; (b): Four-lane closed systems; (c): Four-lane open systems 
in which the right lane has two open ends and the other lanes have PBC. In each panel (a), (b) and (c) 
the figures from left to right correspond to the asymmetric model, the hybrid model and the symmetric model, 
respectively. The red lanes are overtaking lanes and the gray lanes are driving lanes.}
\label{fig:lanes}
\end{figure}

Lane-changing rules in two-lane traffic are in general divided into two
categories: symmetric and asymmetric~\cite{Rickert,Chowdhury}.  This
classification can be generalized to models with more than two lanes.  For
example, a multilane system consisting entirely of driving lanes corresponds to
a symmetric lane-changing model.  An asymmetric lane-changing model in our
program can be made up with a driving lane for the far right lane and
overtaking lanes for all other lanes, or equivalently, it is constructed
entirely with overtaking lanes (see Fig.~\ref{fig:lanes}).  Here we also
consider a case (a hybrid model) in which only the far left lane is the
overtaking lane while the other lanes are driving lanes; this simple
generalization, which differs from an asymmetric model when more than two lanes
are considered, can be regarded as one  minimal model that mimics highway rules
implemented mainly outside of Europe, such as in the U.S. or in many countries
of the Asia-Pacific region.  We are not aware of any previous studies on the
same hybrid model as we consider here.

Below we discuss our results for closed systems with periodic boundary conditions and
for open systems with one open lane separately in two subsections.

\subsection{Closed systems}
\label{sec:closed}

\begin{figure}
\centering
\centerline{\includegraphics[width=\columnwidth]{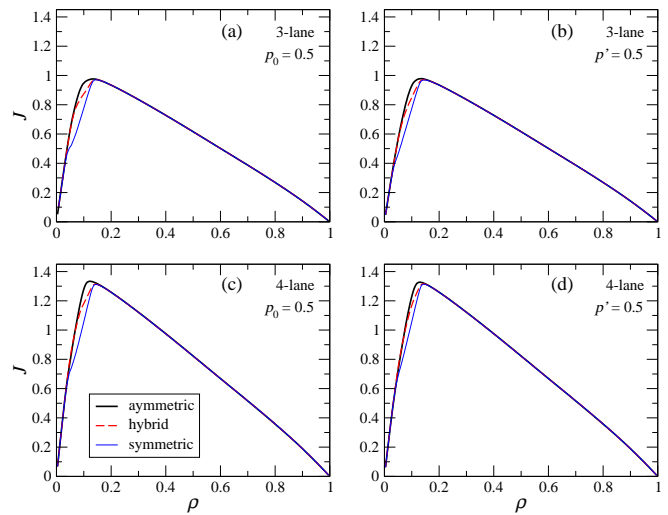}} 
\caption{Fundamental diagrams for three-lane traffic (upper panel (a), (b))  and 
four-lane traffic (lower panel (c), (d)) on a road of length $L=1024$.
The black, red and blue curves indicate data for the asymmetric, hybrid and symmetric lane change models,
respectively. Two types of randomization in braking are considered:
VDR type defined in Eq.~(\ref{eq:VDR}) with $p_0=0.5$ (left panel) and a velocity-independent braking
probability $p'=0.5$ (right panel). The traffic flows for the three
lane change models differ in the free-flow region, in which the flow for
the asymmetric model is the highest, showing the advantage of this model over the other two models.}
\label{fig:fd_closed}
\end{figure}

First we consider closed systems with two types of vehicles
characterized by two different maximum forward velocities
$v^s_{\rm max} = 3 (\rm{cells/time\; step})$ and $v^f_{\rm max} = 5$, 
in which $25\%$ of the vehicles are of slow type. 
We focus on the case with velocity-dependent stochastic braking probabilities $p$:
\be
     p(v)=\begin{cases}
	0  & {\rm for}\;v=v^f_{\rm max}\,,\\
	p_0  & {\rm for}\;v<v^f_{\rm max}\,.	
     \end{cases}
\label{eq:VDR}
\ee
This choice of $p(v)$ corresponds to the so-called ''cruise-control limit'' in which 
fast vehicles at maximum allowed speed move deterministically~\cite{Cruise}.
In the simulations performed, each density value for a system of length $L=1024$ 
was simulated using at least $T=50000$ time steps
and the results were recorded after the first $10000$ steps. 
In addition, each data point is averaged over at least 100 samples.    

\begin{figure}
\centering
\centerline{\includegraphics[width=\columnwidth]{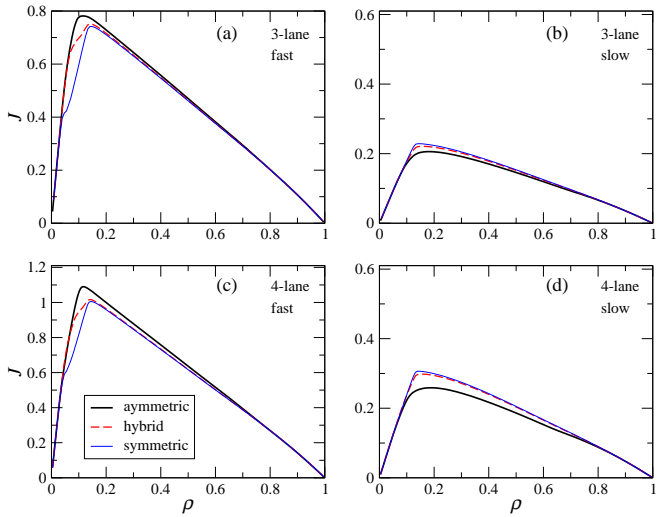}} 
\caption{Flow-density diagrams for fast vehicles (subfigures (a), (c)) and slow vehicles
(subfigures (b), (d))
in three different lane change models with velocity-dependent stochastic braking: $p_0=0.5$ for
$v<v^f_{\rm max}$. The traffic flow of fast vehicles in the asymmetric model is  overall highest,
while the flow of slow vehicles in this model is lower than the other two models.}
\label{fig:fd_types}
\end{figure}

\begin{figure}
\centering
\centerline{\includegraphics[width=\columnwidth]{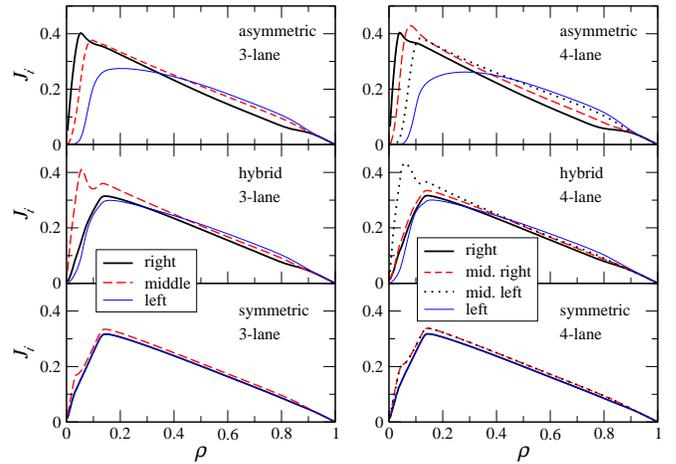}} 
\caption{Flow-density diagrams for each lane in three-lane (left panel)  and four-lane (right panel) 
traffic with the same simulation setup as in Fig.~\ref{fig:fd_types}.}
\label{fig:fd_lanes}
\end{figure}

\begin{figure}
\centering
\centerline{\includegraphics[width=\columnwidth]{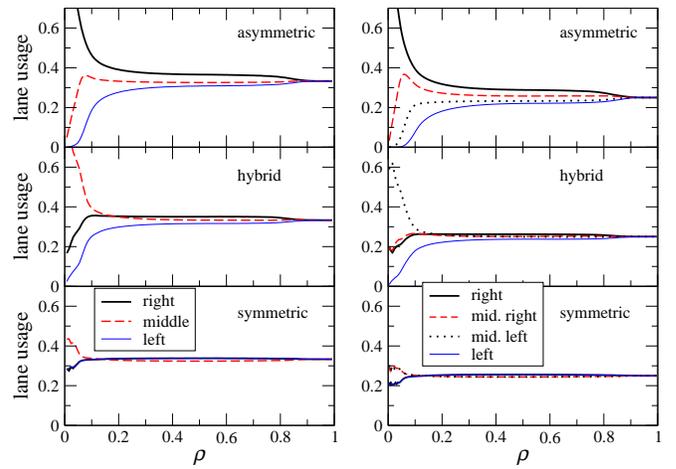}} 
\caption{Lane usage for each lane in three-lane (left panel) and four-lane (right panel)
traffic with the same simulation setup as in Fig.~\ref{fig:fd_types}.}
\label{fig:lu_closed}
\end{figure}

\begin{figure}
\centering
\centerline{\includegraphics[width=0.75\columnwidth]{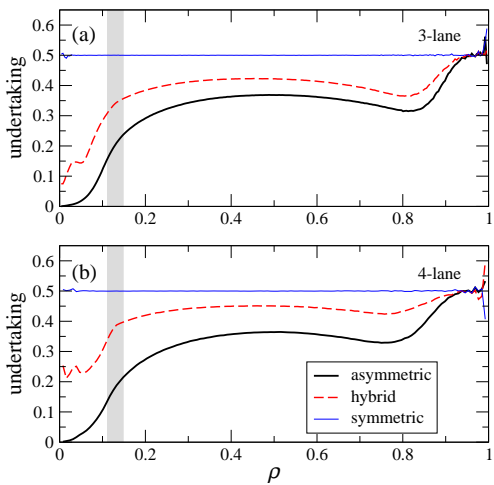}} 
\caption{The fraction of undertaking in three-lane (a) and four-lane (b)
traffic with the same simulation setup as in Fig.~\ref{fig:fd_types}. 
The shaded area marks the densities of maximum flows
in Fig.~\ref{fig:fd_closed}(a) and (c), i.e. the boundaries between a free-flow region
and a jammed phase.}
\label{fig:undertaking}
\end{figure}

Fig.~\ref{fig:fd_closed} shows the fundamental diagrams for
flow($J$)-density($\rho$) relations in three- and four-lane traffic with three
types of lane-changing rules, where results both for the VDR case with
$p_0=0.5$ (see Eq.~(\ref{eq:VDR})) and for the case with a velocity-independent
braking probability $p(v)=p',\;\forall v$ are included.   In all these cases,
with increasing density one finds a transition from a free-flow region at low
density into a jammed region at high density, separated by the peak of traffic
flow.  The diagrams suggest that different lane-changing rules have a stronger
influence on the dynamics in the free-flow phase (close to the maximum of the
flow) than in the jammed region; the asymmetric model, showing a higher
flow in the free-flow phase,  is the most efficient among three
different lane-changing rules while the symmetric model is the least efficient. 
The advantage of the asymmetric model over the other two models with respect
to traffic flow is mainly contributed by the fast vehicles, as we can see in
Fig.~\ref{fig:fd_types}. Slow vehicles, on the other hand, have lower average speed (i.e. 
smaller traffic flow at a given density) in the  asymmetric model than the other two models.

In our program various quantities are measured also with respect to each lane.
For example, Fig.~\ref{fig:fd_lanes} shows the flow-density relations for the individual lanes
in three-lane and four-lane traffic for the VDR case with $p_0=0.5$.
We notice that there are qualitative changes in the flow-density relations for
certain lanes, such as the middle lane of the hybrid model, in which 
an inflection point appears on the right side of the flow maximum. 
The similar feature has been discussed in previous studies on
metastable states in one-lane NaSch model with velocity dependent randomization~\cite{VDR,VDR1}. 

Another interesting quantity for understanding effects of different lane-changing rules
is the lane usage (defined as $N_i/N$ for the $i$-th lane),
shown as a function of density in Fig.~\ref{fig:lu_closed}.
We observe large occupancy of the right lane
in the low-density regime (the free-flow region) of the asymmetric model and the right-lane usage
remains higher than the other lanes when decreasing with growing density approaching $\rho\approx 0.9$, 
which reflects the Keep Right Unless Overtaking rule.   
In the hybrid model, 
the occupancy of the middle lane (or in the middle left lane in
four-lane traffic) dominates at low densities, which is contributed by lane changes from the right side
and also, in particular, from the leftmost passing lane as required by the rules;
interestingly, there is a lane-usage inversion between two ''slow lanes'' of the three-lane model when the density 
becomes larger than $\rho\approx 0.2$.
Unlike the other models, the lane usage in  the symmetric model is evenly distributed over 
all lanes except for a small enhancement 
in the middle lane(s) in the low-density limit.
From the lane usage characteristic along with the flow-density diagrams of 
the individual lanes (Fig.~\ref{fig:fd_lanes}),
we summarize two observations as follows:
(i) non-concavity occurs in the flow-density relation of the lane which exhibits a sharp decay of usage;
(ii) the associated densities of the non-concavity region are those where the sharp decay of
lane usage takes places.

Note that although the strategy of the asymmetric model implies a Keep Right
Unless Overtaking rule, using the lane-changing criteria described in {\bf
LC1}-{\bf LC4} will not avoid ''undertaking'' (i.e. passing on the right) in
free traffic flow, which is prohibited by driving regulations in some
countries.  There are non-vanishing occurrences of undertaking for all
$\rho>0$, as shown in Fig.~\ref{fig:undertaking}. Nevertheless, the fraction of
undertaking in the free-flow phase of the  asymmetric model is considerably
smaller than the fraction in the other two models.  The increasing undertaking
frequency at high densities reflects more symmetric lane usage in congested
traffic.

\subsection{Open systems}
\label{sec:open}

\begin{figure}
\centering
\centerline{\includegraphics[width=\columnwidth]{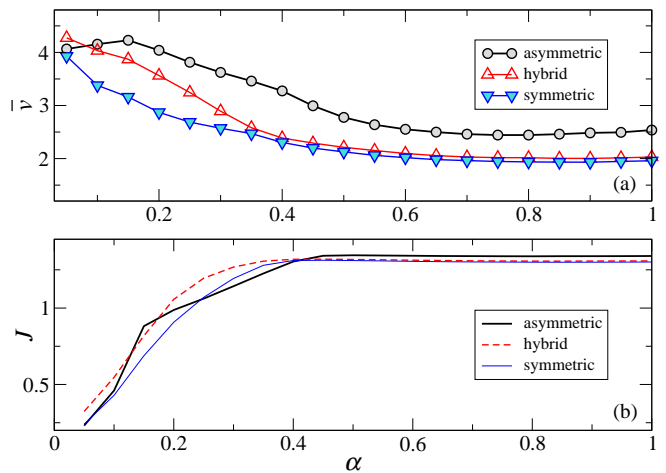}} 
\caption{Average velocities (a) and flows (b) plotted against the entry rate in four-lane open systems.
Three different lane-changing rules with velocity-dependent stochastic braking: $p_0=0.5$ for
$v<v^f_{\rm max}$ are considered.}
\label{fig:v_open}
\end{figure}

\begin{figure}
\centering
\centerline{\includegraphics[width=\columnwidth]{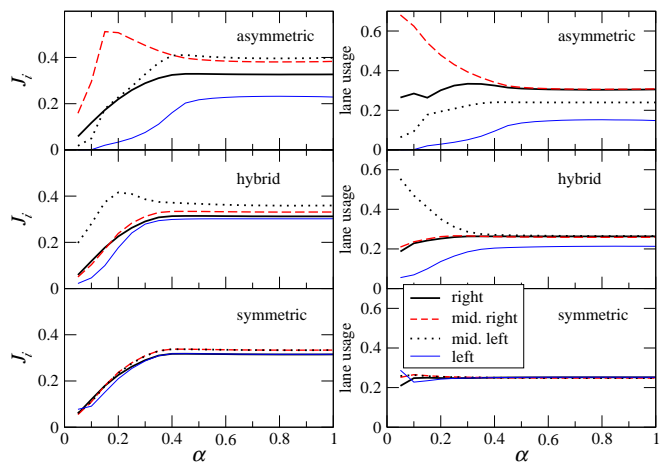}} 
\caption{Flows (left panel) and lane usage (right panel) of the individual lanes
in open systems with four lanes.}
\label{fig:fd_lanes_open}
\end{figure}

\begin{figure*}
\centering
\centerline{\includegraphics[width=8.8cm]{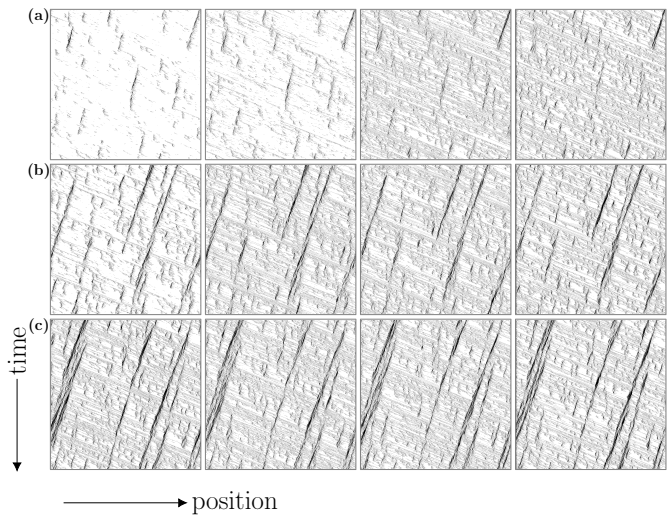}} 
\caption{Space-time diagrams of four-lane open systems using $\alpha=0.95$ and $L=1024$.
Panels (a), (b) and (c) are models with the asymmetric, hybrid and symmetric
lane changing rules, respectively.
From left to right: the left lane, the middle left lane, the middle right lane and the right lane with
two open ends.
In each model, 25\% of vehicles entering the system are of slow type
with a maximum allowed speed $v^s_{\rm max}=3$ and 75\% of vehicles are of fast type
with $v^f_{\rm max}=5$.
In the figure, a black pixel represents a vehicle at a speed $v<5$ and
white pixels are vacancies or vehicles at the highest speed $v=5$.
The data were collected over the last 1024 time steps in total $10^5$ steps.}
\label{fig:space_time}
\end{figure*}

Now we turn to open systems. Here we consider systems which consist of a
multilane part with periodic boundary conditions and a lane on the rightmost side with
open ends, serving as on- and off-ramps (see Fig.~\ref{fig:lanes}(c)).
It has been known for single-lane NaSch models that rules for injection and removal of 
vehicles have significant impacts on the traffic flow~\cite{Open_NaSch,Open_VDR,Open_Ma,Reentrance}.
Here for our multilane model we focus on the following strategies for vehicle injection and removal,
that are incorporated into the single-lane updates and lane changes at each time step: 
\begin{enumerate}
\item {\it Injection strategy:} With probability $\alpha$ a vehicle with initial velocity $v=1$ is inserted into site $j=0$ if the site is not occupied. 
\item {\it Removal strategy:} If a vehicle moves out of the right lane from the open end
at $j=L-1$ by the  NaSch rules ({\bf A1}-{\bf A4}), we remove the vehicle with probability $\beta=1$.
\end{enumerate}
The other parameters used for the simulation, such as
types of vehicles and braking probabilities, are the same to the choices used for the closed systems;
in particular, a VDR rule in the cruise-control limit (Eq.~(\ref{eq:VDR})) with $p_0=0.5$
for stochastic braking is applied, and two different values of
the maximum allowed speed: $v^s_{\rm max} = 3$ and $v^f_{\rm max} = 5$
are assigned to vehicles that enter the system, 
in which $25\%$ of the vehicles are of slow type.
We obtained observables at different entry rates ($\alpha$) and used more than $10^5$ time steps 
for each entry rate.     
The three types of lane-changing rules (asymmetric, hybrid and symmetric)
discussed above are considered here too. 
For the asymmetric model,
we set a default lane left adjacent to the open lane so that 
the criteria for changing back to the open lane are the same
in all three models.

We graph average velocities and flows as functions of $\alpha$ for three
different lane-changing models with four lanes in Fig.~\ref{fig:v_open}.  
In comparison, the velocity in the asymmetric model is overall higher,
showing the advantage of the asymmetric lane-changing rules.
All traffic flows  saturate to constant values at high entry rates, where
the flow of the asymmetric model is slightly larger than the other two models.  

To analyze traffic behavior in each lane, in Fig.~\ref{fig:fd_lanes_open} we
show simulation data for traffic flow and lane usage in the individual lanes,
plotted against the entry rate.  We observe that the flows in the lanes with
high usage fraction (e.g. the middle right lane in the asymmetric model and the
middle left lane in the hybrid model) exhibit non-monotonic behavior before
they converge to constant values at larger $\alpha$; this is similar to what
one observes in the flow-density diagrams of the closed systems shown in
Fig.~\ref{fig:fd_lanes}.    

As a visual demonstration of the effects of lane-changing rules, typical
space-time diagrams for the three different rules in the phase with
$\alpha=0.95$ are shown in  Fig.~\ref{fig:space_time}.  The diagram for the
asymmetric model shows small fluctuations in the two right lanes and low usage
of the leftmost lane, while the plots for the hybrid and symmetric
lane-changing rules exhibit traffic jams that persist for a long time in all
lanes.

\section{Summary and outlook}
\label{sec:summary}

We have studied traffic flows on multilane highways with three different
combinations of driving lanes and overtaking lanes. Lane-changing rules
distinguish between a driving lane and an overtaking lane in the way that in a
(middle) driving lane one makes lane changes to the left and to the right in
symmetric manner, while in an overtaking lane vehicles obey asymmetric rules
for lane changes.  Using lane-changing criteria based on look-ahead distances
we simulated three-lane and four-lane highway traffic in closed systems with
periodic boundary conditions as well as in open systems with on- and off-ramps.
Our results show that for heterogeneous traffic  the asymmetric model with the
rightmost lane as a driving lane and all the other lanes as overtaking lanes, which
mimics the Keep Right Unless Overtaking rule, is more efficient than the other
models in which lane-usage is almost equally distributed in all lanes.

We have developed an extensible software package for our study and make it
publicly available for further applications.  In addition to the observables
considered in this paper, the code covers measurements of the order parameter,
correlations, and relaxation time as defined in Ref.~\cite{Relaxation,Jamming},
which makes it also useful for the study of jamming transitions and dynamic
phase transitions in related
models~\cite{Autocorrelations,Glass,Finite,Absorbing}.

\begin{acknowledgements}
The authors acknowledge support from the 
Ministry of Science and Technology (MOST) of Taiwan under Grants No. 104-2112-M-004-002
and 106-2112-M-004-001.
\end{acknowledgements}

\section*{Author contribution statement}
Y.C.L. devised the project. 
T.Z. developed the software package.
Both T.Z. and Y.C.L performed the numerical simulations. 
Y.C.L. wrote the manuscript.

\end{document}